\documentclass[conference]{IEEEtran}
\IEEEoverridecommandlockouts
\usepackage{booktabs} 
\usepackage{units}
\usepackage{color}
\usepackage{subfloat}
\usepackage{tabularx}
\usepackage{graphicx}
\usepackage{epstopdf}
\usepackage{authblk}
\usepackage{amsmath}
\usepackage{amsthm}
\usepackage{figsize}
\usepackage{subcaption}
\usepackage{enumerate}
\usepackage{epstopdf}

\usepackage{algorithm}
\usepackage{algpseudocode}
\usepackage[table,xcdraw]{xcolor}
\usepackage[figuresright]{rotating}
\usepackage{amssymb}
\usepackage{multicol}
\usepackage{url}
\usepackage{graphicx}
\usepackage{epstopdf}

\title{\LARGE \bf
Uncertainty Quantification in Alzheimer's Disease Progression Modeling
}

\author{ 
        \parbox{3 in}{ \centering Wael Mobeirek, Shirley Mao \\
         University of Illinois at Urbana-Champaign\\
         {\tt\small \{wmobei2, smao10\}@illinois.edu}}

}

\begin{document}

\maketitle
\thispagestyle{plain}
\pagestyle{plain}

\begin{abstract}
With the increasing number of patients diagnosed with Alzheimer's Disease, prognosis models have the potential to aid in early disease detection. However, current approaches raise dependability concerns as they do not account for uncertainty. In this work, we compare the performance of Monte Carlo Dropout, Variational Inference, Markov Chain Monte Carlo, and Ensemble Learning trained on 512 patients to predict 4-year cognitive score trajectories with confidence bounds. We show that MC Dropout and MCMC are able to produce well-calibrated, and accurate predictions under noisy training data.  
\end{abstract}

\section{Introduction}
\label{sec:intro}
Alzheimer’s disease (AD) is a progressive neurodegenerative disease that causes the loss of important neurological functions, including memory. It is the sixth leading cause of death worldwide \cite{adfacts2019}. In America, there are currently 5.8 million patients diagnosed with AD and it is expected to grow to 13.8 million by 2050. In 2019, the total payments for AD-related care was valued at \$290 billion and this is a significant burden on the economy \cite{adfacts2019}. Currently, there is no cure for AD. However, if AD is diagnosed early, it is possible to manage the symptoms to minimize disturbances in daily activities \cite{satone2020predicting}.

In recent years, many studies focused on identifying important biomarkers that can predict the disease’s progression and manifestation. These biomarkers play an important role in developing effective patient-based treatments that aim to slow disease progression \cite{tabarestani2018profile}. In order to tailor such interventions, medical professionals require an accurate estimation of cognitive trajectories and disease manifestation. 

Recent advancements in machine learning led to reasonably successful attempts for utilizing neuroimaging and cross-sectional data for AD diagnosis and prognosis \cite{liu2020multi}. However, they do not model the progression of the disease over time. Other approaches leverage time-series data to create disease progression models (DPMs) to predict the progression of the clinical status in AD, which are categorized as cognitively normal (CN), mild cognitive impairment (MCI), and dementia \cite{el2020multimodal}. The existing approaches do not account for uncertainty in the data and the model, which raises concern for medical experts and patients regarding the reliability and dependability of the methods.

In this work, we aim to estimate uncertainty in AD’s prognosis as an additional validation of the forecasts of cognitive scores to provide trustworthy and reliable predictions. We compared four models for predicting the future trajectories of the Mini-Mental State Exam (MMSE) score: Monte Carlo Dropout (MC Dropout), Variational Inference (VI), Markov Chain Monte Carlo (MCMC), and Ensemble Learning.

The contributions of this work are as follows:
\begin{itemize}
 \item Proposal of prognosis models for AD progression that produce confidence bounds.
 \item A comparative study between various uncertainty quantification (UQ) methods for ADPM.
\end{itemize}


\begin{figure*}[t!]
\centering
\includegraphics[width=0.9\textwidth]{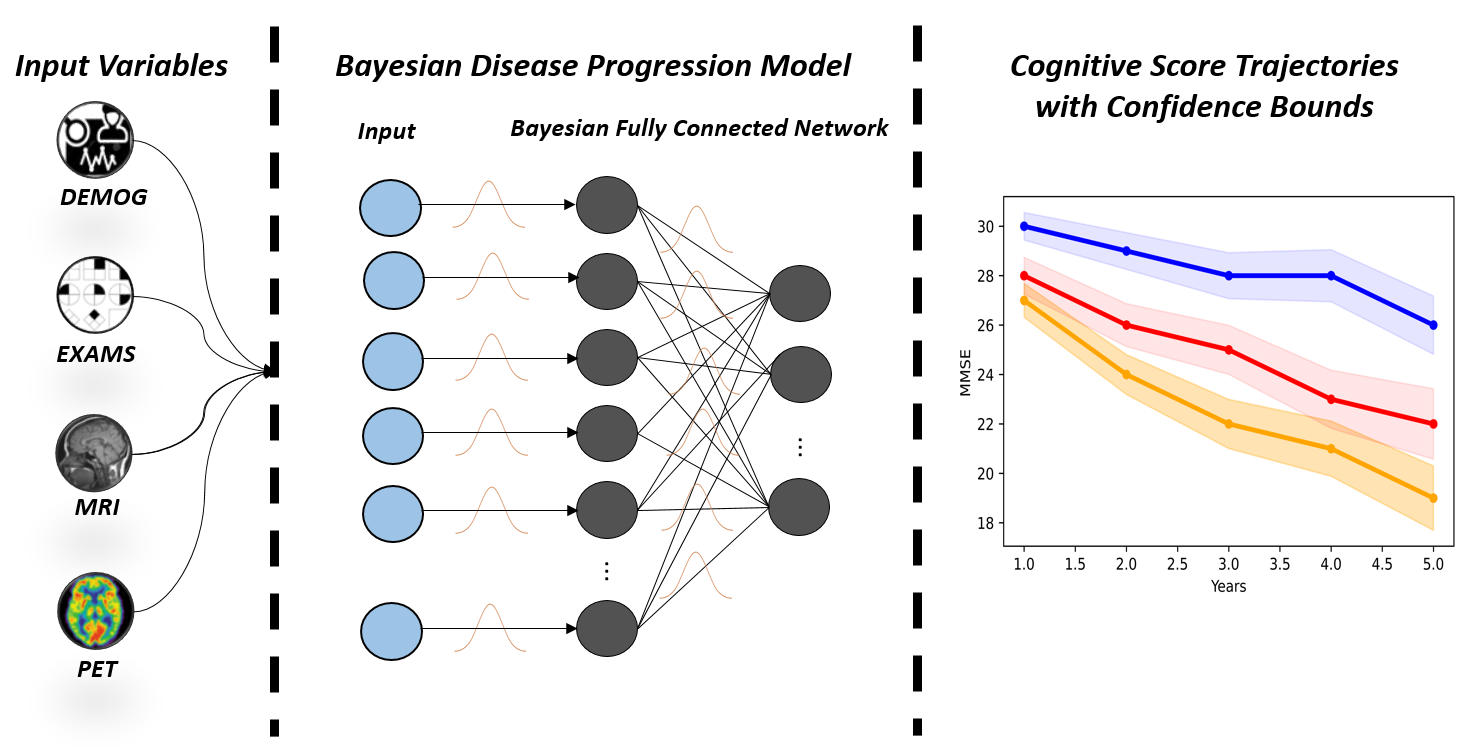}
\caption{\label{fig:HighLevel} A high level representation of the input variables and the proposed Bayesian Setup} 
\end{figure*}

\section{Background}
\subsection{Alzheimer’s Disease Progression
Models}
Research in ADPMs has been emerging in recent years. Nyugen et al. \cite{nguyen2020predicting} used minimalRNN to jointly deal with missing values and predict future cognitive trajectories. They modeled the patient’s individual states as a latent vector representation and used this representation to impute missing values. Jung et al. \cite{jung2021deep} built on the work in \cite{nguyen2020predicting}, and proposed a deep neural network model to jointly impute missing values in the input data, forecast biomarkers measurements of the patients, estimate future trajectories of cognitive scores, and predict the clinical status of the patient at future years. In a recent work, Saboo et al. \cite{saboo2021reinforcement} combined differential equations and reinforcement learning with domain knowledge to predict 10-year cognition trajectories during AD’s progression. The model not only achieved comparable performance to state-of-the-art approaches, but also provided insightful explanations of the compensatory process in AD.

Previous approaches focused exclusively on improving the error of cognitive decline predictions without quantifying uncertainty in the model’s decision. Fruehwirt et al. \cite{fruehwirt2018bayesian} are the first to study AD’s progression modeling in a probabilistic setup. They proposed a Bayesian neural network to obtain a predictive distribution and impose uncertainty bounds on the predicted trajectories of cognitive scores using quantitative EEG measurements as input. However, \cite{fruehwirt2018bayesian} only studied uncertainty from measurements at baseline only and did not consider multiple important biomarkers that are directly related to AD’s progression.

We aim to address these limitations by modeling uncertainty in AD’s progression by using MC Dropout, Variational Inference, MCMC, and Emsemble Learning for 4 years into the future.
\subsection{Uncertainty Quantification Methods}
Bayesian methods provide a mechanism for uncertainty quantification \cite{abdar2021}. The Bayesian posterior can be written as follows:
\begin{equation}
p\left(w|x,y\right)= \frac{p\left(y, w |x\right)}{p\left(y|x\right)}
\end{equation}
A Bayesian model is trained to estimate the posterior probability $P(w|\mathcal{D})$ where $P(w)$ is the prior distribution of the weights and $\mathcal{D}$ is the training dataset. At inference time, the likelihood for the given parameters can be calculated as following:
\begin{equation}
p\left({y}^{*} \mid {X}^{*}, \mathcal{D}\right)=\int_{w} p\left({y}^{*} \mid {X}^{*}, w\right) p(w \mid \mathcal{D}) dw
\end{equation}
where ${y*}$ is the sampled output and ${X*}$ is the test sample.

However, the posterior $P(\theta|\mathcal{D})$ is intractable, but it is possible to approximate a distribution ${q_{\theta}}\left({W}\mid {\theta}\right)$  that is close to the posterior distribution obtained by the model. We will discuss four Bayesian approximation methods: Monte Carlo Dropout, Variational Inference, Monte Carlo Markov Chain, and Ensemble Learning. 

\subsubsection{Monte Carlo (MC) Dropout} approximates the posterior by applying Dropout during inference time.
Dropout is a popular regularization technique used during training where neurons are randomly “shut off”. By applying Dropout at inference time, Gal et al \cite{galmcd2015} showed that the posterior can be approximated without sacrificing computational complexity nor accuracy.
\subsubsection{Variational Inference} approximates the posterior over the weights by minimizing the Kullback–Leibler (KL) divergence between the two distributions $P({W}|{D})$ and   ${q_{\theta}}\left({W}\mid {\theta}\right)$:
\begin{equation} 
 D_{K L}\left(q_{\theta} \mid P\right)=\int_{{W}} q_{\theta}\left({W} \mid \theta \right) \log \left(\frac{q_{\theta}\left({W} \mid \theta \right)}{P\left({W} \mid D\right)}\right) d {W}
\end{equation}
The loss can be simplified to the Evidence Lower Bound (ELBO):
\begin{equation}
\begin{split}
&D_{K L}\left(q_{\theta} \left({W} \mid \theta \right) \mid P \left({W} \mid \theta \right)\right) \\
&= D_{K L}\left(q_{\theta} \left({W} \mid \theta \right) \mid P \left({W} \right)\right) 
-\mathbb{E}_{q({w} \mid \boldsymbol{\theta})} \log P(\mathcal{D} \mid {w})
\end{split}
\end{equation}
While KL-divergence benefits from the stochastic optimization methods, the results heavily depend on the starting point \cite{swiatkowski2020}
\subsubsection{Markov Chain Monte Carlo (MCMC)} approximate the posterior of a BNN’s weights by randomly drawing values from the distribution and applying a stochastic transition to them. After many repetitions, the outcome eventually converges to the distribution of the posterior. However, the sufficient number of iterations is unknown \cite{kupinski2003}.
\subsubsection{Ensemble Learning} combines multiple models to obtain uncertainty estimates. Independent models that are initialized with different models are trained on the same dataset, then are used for sampling at test time \cite{malinin2019}. Ensembles have shown that they are able to provide robust predictions and robust uncertainty estimates \cite{abdar2021} 
\section{Data \& Approach}
\subsection{Data}
\subsubsection{Data Summary}
\begin{figure*}[t!]
\centering
\includegraphics[width=0.9\textwidth]{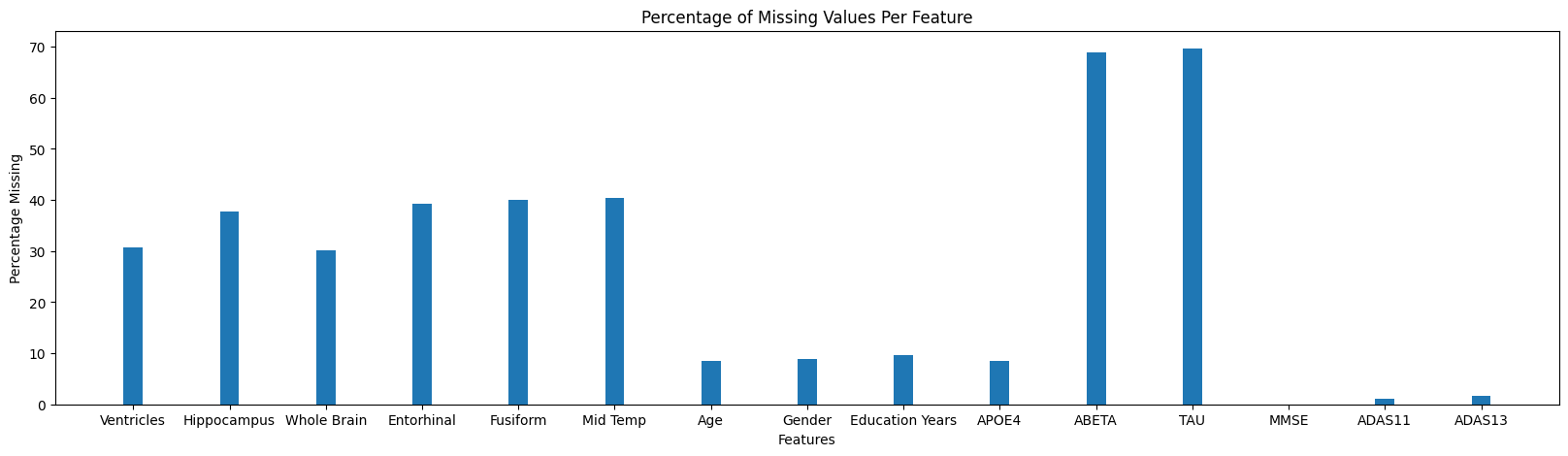}
\caption{\label{fig:MisingValues} The percentage of missing value per feature in the filtered dataset used for imputation.} 
\end{figure*}

We compare the performance of the proposed approach on the Alzheimer’s Disease Neuroimaging Initiative (ADNI) database’s TADPOLE longitudinal data \cite{tadpole}, which includes imaging biomarkers, clinical measurements, and cognitive test scores from 2155 patients aged between 54.4 and 91.4 years with 13,915 visits at 26 different time points between 2003 and 2019. The dataset is noisy due to the inherent diversity in the AD population and has many missing values.

\subsubsection{Feature Selection}
We considered six volumetric features of the ventricles, hippocampus, whole-brain, entorhinal cortex, fusiform gyrus, and middle temporal gyrus as imaging features. These values are normalized with respect to each patient’s intracranial volume to account for variability in the brain size. Additionally, we included age, gender, years of education as demographic features, the presence of the APO4 gene as a genetic feature, and amyloid-beta level, and tau protein level as lab examination features. For cognitive scores, three exams were considered, which are the Mini-Mental State Exam (MMSE), The Alzheimer's Disease Assessment Scale-Cognitive Subscale 11 (ADAS-Cog 11), and The Alzheimer's Disease Assessment Scale-Cognitive Subscale 13 (ADAS-Cog 13). The ranges of the scores of these exams are [0-30], [0-70], and [0-85] respectively, where 0 is the worst cognitive performance. We focus on MMSE as our target cognitive score for prediction. 

\subsubsection{Data processing \& Missing Value Imputation}
To process the data, we filtered out patients with missing MMSE scores and no five yearly consecutive visits. This reduced the total number of patients to 512. Then, we leveraged Jung et al \cite{jung2021deep}'s work in imputing the missing values. A summary of the percentage of missing values per feature is shown in Figure \ref{fig:MisingValues}. Finally, we perform z-score normalization and split the data 90/10/10 for training, validation, and testing respectively.

\subsection{Approach}
\begin{figure*}[t!]
    \centering
    \includegraphics[width=0.9\textwidth]{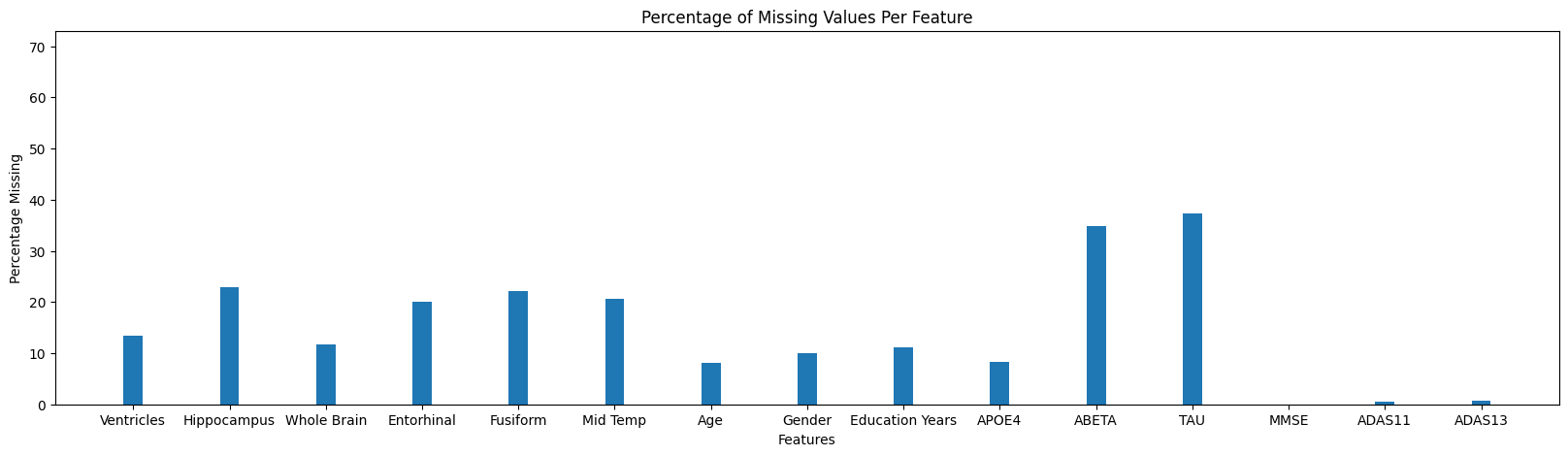}
    \caption{\label{fig:TrainingMisingValues} The percentage of missing value per feature used for training.} 
\end{figure*}
We propose training the methods on the processed ADNI dataset \cite{tadpole} using the baseline year of measurements for training and evaluate models on predicting the MMSE score for years 1, 2, 3 and 4. A summary of the percentage of missing values per feature that are imputed in the baseline year for training is shown in Figure \ref{fig:TrainingMisingValues}. To avoid the computational complexity of training MCMC and Ensemble, we simplify the architecture of all the methods to include a fully-connected neural network with a single hidden layer. Each model has 4 output neurons, 1 per future year. To select the priors, we follow Foong et al's \cite{foong2019}  recommendation of choosing $\mathcal{N}(0, \sigma^2/N_{in})$ for all the weights, with  $\sigma^2$ as a constant and $N_{in}$ as the number of features. For hyperparameter optimization, we used grid search in the case of MC Dropout, VI, and Ensemble Learning. For MCMC, we performed manual experimentation to optimize the hyperparameters. We perform Monte Carlo Sampling to generate 500 samples per cognitive trajectory.

\subsection{Evaluation Metrics}
To evaluate the predictive performance and the quality of the uncertainty estimates, we propose the following metrics.

First, we propose using the root mean square error (RMSE) to evaluate the predictive performance defined as follows: 

\begin{equation}
\operatorname{RMSE}=\sqrt{\frac{\sum_{i=1}^{N}\left(x_{i}-\hat{x}_{i}\right)^{2}}{N}}
\end{equation}

where $x_{i}$ is the actual observation, $\hat{x}_{i}$ is the predicted value, and $N$ is the number of test points.

To evaluate the confidence of the model, we propose using the predictive variance ($\sigma^{2}$) is defined as follows:
\begin{equation}
\sigma^{2}=\frac{\sum_{i=1}^{N}\left(x_{i}-\bar{x}_{i}\right)^{2}}{N}
\end{equation}

where $x_{i}$ is the predicted value, $\bar{x}_{i}$ is the mean of all of the predicted values, and $N$ is the number of test points.

To evaluate the quality of the uncertainty estimates,  we propose using relative accuracy and the miscalibration area. Relative accuracy measure the percentage of true values that fall within the confidence interval of the model as follows:
\begin{equation}
accuracy_{relative }=\frac{\sum_{i=1}^N I_{\left\{C I_{\text {Low }_i} \leq x_i \leq C I_{\text {High }}\right\}}}{N}
\end{equation}

Calibration, on the other hand, states that the predicted values should fall in the 90\% confidence interval approximately
90\%  of the time . The calibration error proposed by Chung et al \cite{chung2021uncertainty} is a convenient numerical score to describe the quality of the calibration defined as follows: 
\begin{equation}
\operatorname{cal}\left(F_1, y_1, \ldots, F_T, y_T\right)=\sum_{j=1}^m w_j \cdot\left(p_j-\hat{p}_j\right)^2
\end{equation}
for a confidence level $0 \leq p_1 \le p_2 \le ... \le p_m \leq 1$ where $F$ is a forcaster model, $y$ is the predicted value, and $w_j$ is a scalar constant set to $1$.  

\subsection{Experimental Setup}
The experimental setup in this work is built using PyTorch
library for MC Dropout, VI, and Ensembles. For MCMC, we use Pyro with JAX running in the backend for computational stability. Pyro's default MCMC technique is Hamiltonian Monte Carlo. To run the experiments, we used a Windows 10 machine with Intel Core i9 9900K for the central processing unit, NVIDIA GeForce 1080 Ti graphics processing unit, and 32 GB of random access memory.

\section{Results}
\subsection{Predictive Performance}
\begin{table}[H]
    \footnotesize
    \begin{tabular}{|l|c|c|c|c|}
    \hline
    Methods &
      \multicolumn{1}{l|}{RMSE} &
      \multicolumn{1}{l|}{Variance} &
      \multicolumn{1}{l|}{\begin{tabular}[c]{@{}l@{}}Relative \\ Accuracy\end{tabular}} &
      \multicolumn{1}{l|}{\begin{tabular}[c]{@{}l@{}}Miscalibration Area\end{tabular}} \\ \hline
    \begin{tabular}[c]{@{}l@{}}Determi-\\ nistic NN\end{tabular}     & 1.679 & -     & -       & -\\ \hline
    MC Dropout                                                       & 1.673 & 1.390  & 75.49\% & 0.12  \\ \hline
    \begin{tabular}[c]{@{}l@{}}Variational \\ Inference\end{tabular} & 1.693 & 0.295 & 49.01\% & 0.30 \\ \hline
    MCMC                                                             & 1.614 & 6.409  & 99.01\%  & 0.16 \\ \hline
    Ensemble                                                         & 1.756 & 0.411 & 28.43\% & 0.35
    \\\hline
    \end{tabular}
    \caption{\label{fig:ResultsTable}Summary of the average results across the years.} 
\end{table}
We used the deterministic neural network as a sanity-check measure and included its RMSE values in the table.
MCMC achieved the smallest RMSE value while Ensemble Learning had the largest error. Variational inference had the smallest variance and MCMC had the largest variance. The highest and lowest relative accuracy is MCMC and Ensemble respectively. The highest and lowest miscallibration area is Ensemble and MC Dropout respectively. All the methods except for MCMC were confident in their prediction.
\\\\Taking all the fields into account, MC Dropout performed the best overall in predicting the 4-year cognitive trajectory.
\subsection{Quality of Uncertainty Estimates}
\label{sec:uqResults}

\begin{figure*}[h]
\centering
\begin{multicols}{2}
    \includegraphics[width=0.7\linewidth]{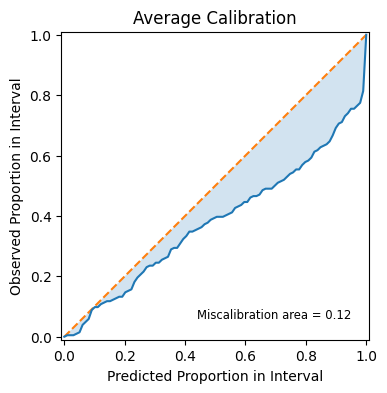}
    \caption{\label{fig:MCDCal} MC Dropout Calibration.} 
    \includegraphics[width=0.7\linewidth]{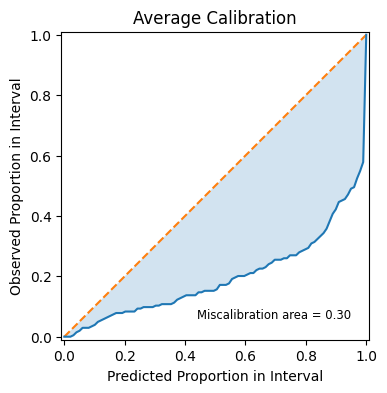}
    \caption{\label{fig:VICal} VI Calibration.} 
    \end{multicols}
\begin{multicols}{2}
    \includegraphics[width=0.7\linewidth]{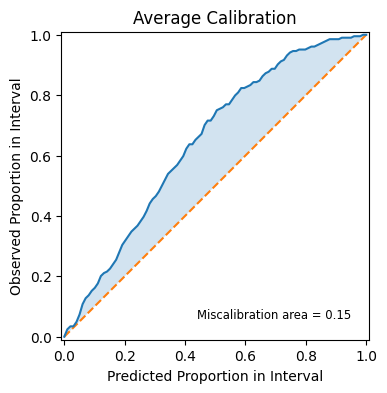}
    \caption{\label{fig:MCMCCal} MCMC Calibration.}
    \includegraphics[width=0.7\linewidth]{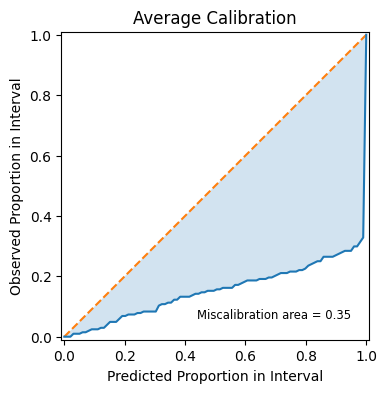}
    \caption{\label{fig:EnsembleCal} Ensemble Calibration.} 
\end{multicols}
\end{figure*}
For figures \ref{fig:MCDCal}, \ref{fig:VICal}, \ref{fig:MCMCCal}, and \ref{fig:EnsembleCal}, the blue curve is the model's calibration and the yellow curve represents perfect calibration (ideal case). The area between the blue and yellow curve is the miscalibration area. The smaller the difference, the closer to the ideal calibration curve the model is. MC Dropout has the smallest miscalibration area while Ensemble has the largest miscalibration area. For MC Dropout, VI, and Ensemble, the model's calibration falls below the perfect calibration. This means these models are very confident with their predictions. MCMC's calibration curve is above the perfect calibration. Hence, it is not very confident.

\begin{figure*}[h]
\begin{multicols}{2}
    \includegraphics[width=\linewidth]{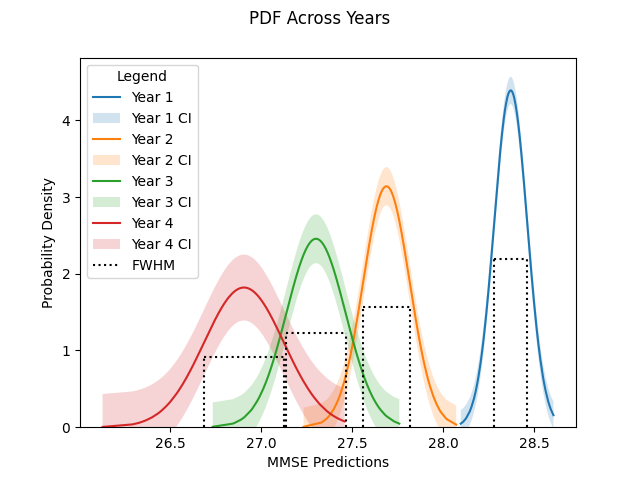}
    \caption{\label{fig:MCDPDF} MC Dropout probability density functions.} 
    \includegraphics[width=\linewidth]{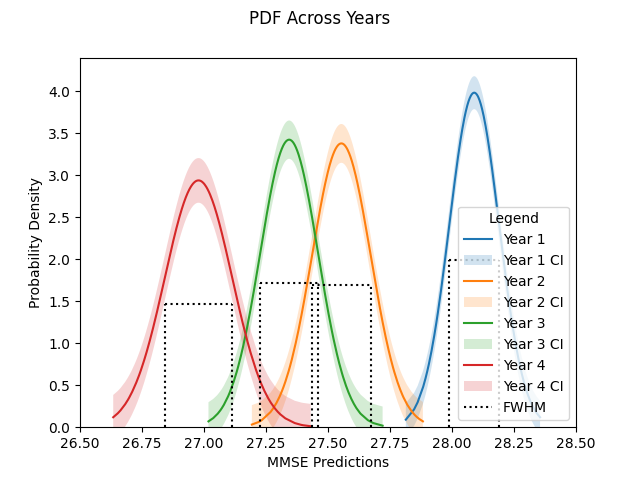}
    \caption{\label{fig:VIPDF} VI probability density functions.} 
    \end{multicols}
\begin{multicols}{2}
    \includegraphics[width=\linewidth]{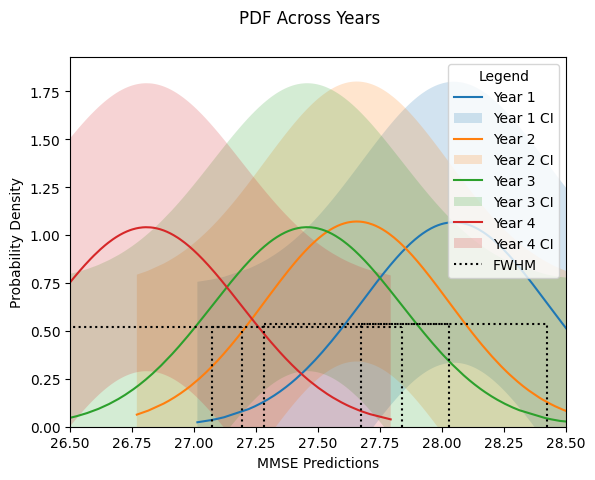}
    \caption{\label{fig:MCMCPDF} MCMC probability density functions.}
    \includegraphics[width=\linewidth]{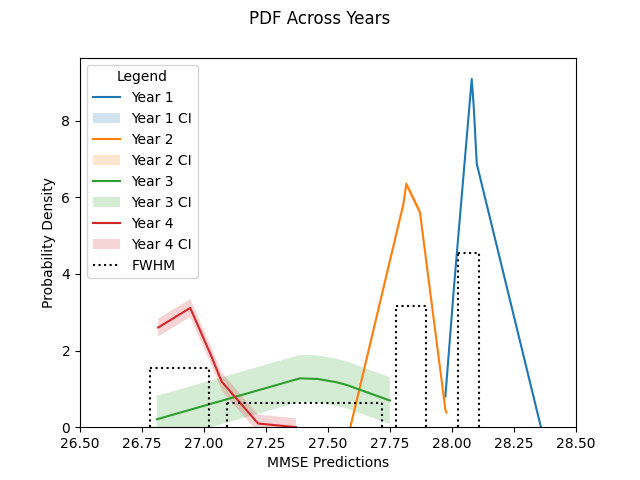}
    \caption{\label{fig:EnsemblePDF} Ensemble probability density functions.} 
\end{multicols}
\end{figure*}
 Figures \ref{fig:MCDPDF}, \ref{fig:VIPDF}, \ref{fig:MCMCPDF}, and \ref{fig:EnsemblePDF} are the plot for the probability density function (PDF) of the 4-year score prediction for each of the methods. A trend we see is the mean MMSE prediction decreases as we attempt to predict further into the future. Additionally the curves reflect the results from Table \ref{fig:ResultsTable} and figures \ref{fig:MCDCal}, \ref{fig:VICal},\ref{fig:MCMCCal}, and \ref{fig:EnsembleCal}. Since MC Dropout, VI, and Ensemble are more confident, the confidence interval is also smaller. On the other hand, MCMC has large confidence interval and thus is not confident.
 \begin{figure*}[h]
\begin{multicols}{2}
    \includegraphics[width=\linewidth]{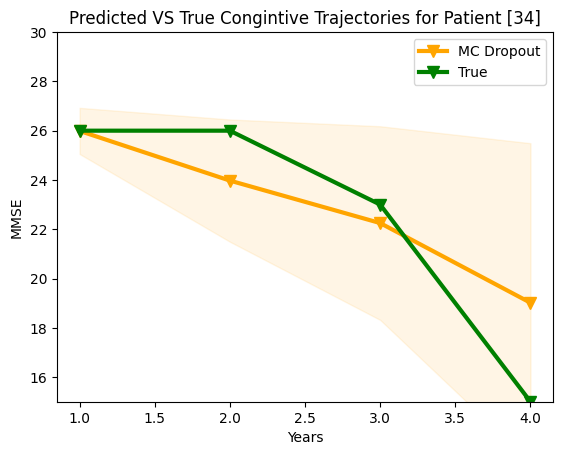}
    \caption{\label{fig:MCDp34} MC Dropout probability density functions.} 
    \includegraphics[width=\linewidth]{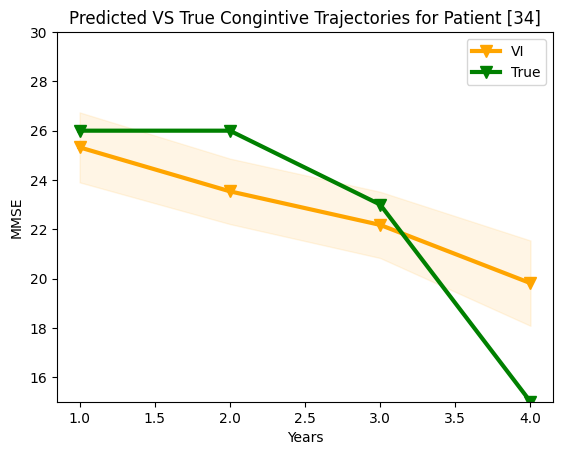}
    \caption{\label{fig:VIp34} VI probability density functions.} 
    \end{multicols}
\begin{multicols}{2}
    \includegraphics[width=\linewidth]{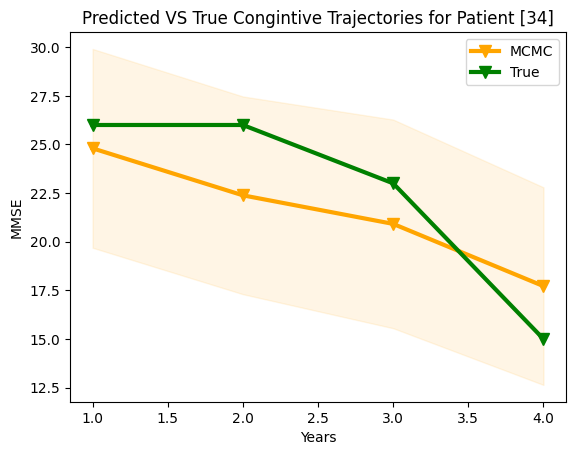}
    \caption{\label{fig:MCMCp34} MCMC probability density functions.}
    \includegraphics[width=\linewidth]{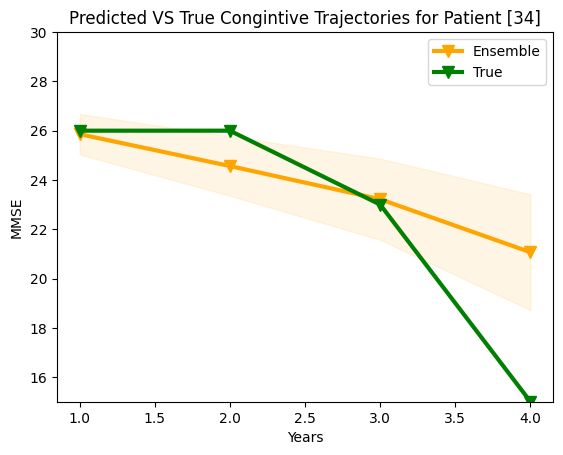}
    \caption{\label{fig:Ensemblep34} Ensemble probability density functions.} 
\end{multicols}
\end{figure*}

Figures \ref{fig:MCDp34}, \ref{fig:VIp34}, \ref{fig:MCMCp34}, and \ref{fig:Ensemblep34} shows the predicted and true cognitive trajectories for the same patient but with the different methods. This further supports the claim that MC Dropout, VI, and Ensemble are more confident due to the tighter bounds for the confidence intervals and MCMC is less confident due to the looser bounds for its confidence interval. Additionally, we can also see that while VI and Ensemble are very confident, the true values are often not within the confidence interval. On the other hand, the majority of true values fall within MC Dropout and MCMC's confidence bounds. This trend is supported by the relative accuracy values in table \ref{fig:ResultsTable}.
\section{Discussion \& Future Directions}
First, we discuss the predictive performance of 4-year cognitive trajectories. Figures \ref{fig:MCDPDF}, \ref{fig:VIPDF}, \ref{fig:MCMCPDF}, and \ref{fig:EnsemblePDF} show that the models were able to capture the decline in cognition as the models, which is expected as the disease progresses. The performance of all models perform competitively as the data imputations improve the predictive ability of the models. Although MC Dropout and MCMC were able to beat the performance of the deterministic neural network, VI and Ensemble were not. We believe that this due to the difficulty of choosing priors in the case of VI. Ensembles had the worst predictive performance due to the random initialization of the independent models combined with the small dataset size as not all of the ensemble models are able to converge optimally. Abdar et al argues that although ensemble usually are able to give robust predictions, their performance suffers in smaller problems \cite{abdar2021}. This is reflected in figure \ref{fig:EnsemblePDF} as the PDFs are sharper compared to the other models, suggesting .  Overall, all methods are able to predict the 4-year cognitive trajectory of patients reasonably well.

Next, we discuss the quality of uncertainty estimates. Ensemble, VI, and MC Dropout were, respectively, overconfident discussed in section \ref{sec:uqResults}. This is expected for MC Dropout and VI as Foong et al \cite{foong2019} has shown that they tend to oversample certain clusters in the dataset while ignoring other clusters causing the posterior approximations produced by these methods to be further away from the exact posterior. Ensemble was the most overconfident method, yet the worst performer, which degrades trust in the model and the uncertainty estimates. This can be explained by the inability of the Mean Sqaure Error as a loss function during training to capture predictive uncertainty as shown by Lakshminarayanan et al \cite{Lakshminarayanan2016}. Instead, the authors recommend using Negative Log-Likelihood (NLL) to obtain well-calibrated results. MCMC was the least confident method, which is comparable to the results obtained by Foong et al \cite{foong2019}. This can be explained by the difficulty in choosing priors that are representative of the data as well as the computational difficulty associated with MCMC as its impossible to determine if the model has converged.

Overall, MC Dropout and MCMC seem to be the most effective UQ technique in producing well-calibrated, highly performant cognitive trajectories. MC Dropout in particular does not require setting priors and it is very efficient. This conclusion can be confirmed by MC Dropout and MCMC having the lowest miscalibration area and the lowest RMSE. Qualitative examination of figures \ref{fig:MCDp34}, \ref{fig:VIp34}, \ref{fig:MCMCp34}, and \ref{fig:Ensemblep34} show that the true trajectory of the patients is captured within the confidence interval by these two methods with close mean prediction. In the future, we would like to explore more Ensemble methods, such as training with NLL as a loss function. Additionally, we would like to use time-series based Bayesian Neural Networks, such as Bayesian Long-Short Term Memory models \cite{hochreiter1997long}, to capture temporal relationships in the dataset. Finally, we would like to study the robustness of these UQ methods against input perturbations to combat against noisy data. 

\newpage
\section{Conclusion}
\label{sec:conc}
This work addresses the problem of uncertainty in Alzheimer's Disease Progression Modeling. We propose models that are able to predict cognitive trajectories of Alzheimer's Disease patients for 4 years into the future while generating confidence bounds for these estimates to improve reliability of these models. The models utilize an imputed dataset with demographic, cognitive, genetic, and imaging features. We also perform a comparative study between four uncertainty quantification techniques: Monte Carlo Dropout, Variational Inference, Monte Carlo Markov Chain, and Ensemble Learning. We show that MC Dropout and Monte Carlo Markov Chain are able to produce reasonably accurate trajectories while providing well-calibrated uncertainty estimates.  



\section*{Acknowledgement}
We thank Mosbah Aouad, Yogatheesan Varatharajah, \& Ravishankar K Iyer for their contributions and feedback on this work.
\bibliographystyle{unsrt}
\bibliography{reference}
\end{document}